\documentclass[aps,prl,showpacs,twocolumn,superscriptaddress]{revtex4}
 \usepackage{amsmath}
\usepackage{graphicx}
\usepackage{epsfig}
\usepackage{subfigure}
\newcommand{\beq}{\begin{equation}}
\newcommand{\eeq}{\end{equation}}
\newcommand{\bea}{\begin{eqnarray}}
\newcommand{\eea}{\end{eqnarray}}
\newcommand{\pdag}{{\phantom{\dagger}}}

\begin{document}
\bibliographystyle{apsrev}
 
\title{Interference as a Probe of Spin Incoherence in Strongly Interacting Quantum Wires  }
\author{M. Kindermann}
\affiliation{ Laboratory of Atomic and Solid State Physics, Cornell
  University, Ithaca, New York 14853  }
\author{P. W. Brouwer}
\affiliation{ Laboratory of Atomic and Solid State Physics, Cornell
  University, Ithaca, New York 14853  }
\author{A. J. Millis}
\affiliation{ Department of Physics, Columbia University, 538 West
  120th Street, New York, New York 10027  }

\date{March 2006 }
\begin{abstract}
 We show that  interference experiments can be used to identify the 
 spin-incoherent regime of strongly interacting one-dimensional
 conductors.  Two qualitative
 signatures of spin-incoherence are found: a strong magnetic
 field dependence of the interference contrast and an anomalous
 scaling of the interference contrast with the applied voltage, with a  
temperature and  magnetic field
dependent scaling exponent. The experiments  distinguish
the spin-incoherent from the spin-polarized regime, and so may be
useful in deciding between alternative explanations proposed for
the anomalous conductance quantization observed in quantum point
contacts and quantum wires at low density.
\end{abstract}
\maketitle

One dimensional conductors are of intense current  interest because they 
generically display strong "Luttinger liquid" interaction effects, such as scaling with nontrivial powerlaws 
\cite{kn:giamarchi2002}. 
Over the last decade
high quality quantum wires have been 
fabricated and shown to display characteristic Luttinger liquid properties \cite{kn:yao1999,kn:auslaender2002}.
A particularly interesting limit is the ``spin-incoherent'' regime \cite{kn:cheianov2004,kn:cheianov2004b,kn:fiete2004,kn:fiete2005,kn:fiete2005b,kn:fiete2006,kn:cheianov2005,kn:kindermann2006}
  in which the energy scale $J$ of  spin excitations
is much less than the temperature $T$. Spin-incoherent behavior is a generic feature of the low
density limit of an interacting one-dimensional electron gas, but it may  also be realized in ultra-thin conductors at 
high electron density   \cite{kn:fogler2005}.   

An important motivation for study of the spin-incoherent regime is provided by the discovery 
of anomalous conductance quantization in 
quantum wires and quantum point
contacts \cite{kn:thomas1996,kn:cronenwett2002,kn:kane1998,kn:reilly2001,kn:thomas2000,kn:biercuk2005}. 
As the electron density is reduced in these systems,
the conductance exhibits a series of plateaus  at  integer multiples of 
$2 e^2/h$ \cite{kn:vanwees1988,kn:wharam1988}. At the lowest densities, however,
an additional  
plateau is observed, at a reduced conductance   value between $0.5 \times 2e^2/h$ and $0.7 \times 2 e^2/h$.  While
conductance quantization at multiples of $2 e^2/h$ can be well
understood within the framework of non-interacting electrons \cite{kn:vanwees1988,kn:wharam1988}  
electron-electron interactions are believed to be responsible for the
reduced conductance at the lowest  plateau. 

Two explanations of the reduced conductance have been proposed:
a spin-polarization of conduction electrons  \cite{kn:reilly2002,kn:meir2002,kn:klironomos2005b} 
and the formation of  a Wigner crystal with  small exchange energy $J$ at low electron
density \cite{kn:matveev2004,kn:matveev2004b}.  Matveev
showed that the spin-incoherent regime at $kT\gg J$ was characterized by 
a conductance plateau with a reduced conductance, consistent with experiments.
Despite their different physical content --- spin-polarization versus spin-incoherence --- 
both models are consistent with a wide range of  experimental
observations, both of   conductance and of  current fluctuations
\cite{kn:roche2004,kn:kindermann2006}.  While a spontaneous spin-polarization can be experimentally detected in  magnetic 
focusing experiments  \cite{kn:rokhinson2006}, the technique cannot easily be applied to longer
wires such as carbon nanotubes  \cite{kn:kane1998,kn:reilly2001,kn:biercuk2005},
where the Lieb-Schulz-Mattis theorem suggests that a ferromagnetic state is unlikely.
A direct probe of the existence of the spin-incoherent regime is therefore needed.

\begin{figure}
\includegraphics[width=7cm]{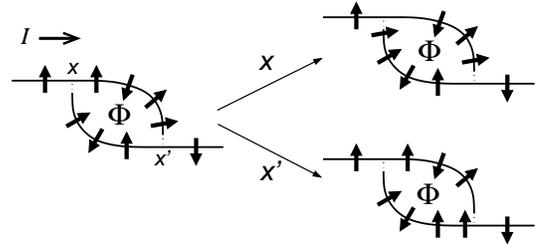}
\caption{ First interference geometry. In spin-polarized conductors  amplitudes for an electron to tunnel from the left to the right wire at the points $x$ and $x'$ interfere. Their phase difference $\varphi=e \Phi/\hbar c$ depends on the flux $\Phi$ through the interference loop.  In the spin-incoherent regime they generically involve different spin configurations that prevent interference. }  \label{fig1} 
\end{figure}

In this Letter we show that quantum interference experiments provide a powerful probe of the presence and properties of spin-incoherence in one
dimensional conductors.  We consider the standard interference  geometry sketched
in Fig \ref{fig1}: a loop enclosing a magnetic flux $\Phi$. We shall be interested
in the interference contrast $C$: the relative change of current $I$ with the phase $\varphi=e \Phi/\hbar c$ 
\begin{equation}
C=\frac{\sqrt{\langle \left(I(\varphi)-\langle I\rangle_\varphi\right)^2 \rangle_\varphi}}{\;\;\;\langle I\rangle_\varphi}, \;\;\;\langle \dots \rangle _\varphi=\int_0^{2\pi}{\frac{d\varphi}{2\pi} \dots} .
\label{cdef}
\end{equation}
Typically the interference contrast decays exponentially with the linear dimension $l$ of the interference 
region provided that $l$ is greater than a phase coherence length $l_\varphi$. In strongly interacting one dimensional
systems  electrons decompose into charge and spin  excitations with generically different velocities $v$ and $v_\sigma$.  The coherence length $l_\varphi^\sigma \sim v_\sigma/T$ for spin excitations may become much smaller
than that of charge excitations, $l_\varphi^\rho \sim v/T$.   In the limit $l_\varphi^\rho \gg l\gg l_\varphi^\sigma$, where charge excitations can move coherently
through the interference region but spin excitations cannot, the interference 
contrast is thus strongly suppressed by interactions in the system.  This is particularly evident in the spin-incoherent regime of  impenetrable electrons ($l_\varphi^\sigma \lesssim a$, where $a$ is  of the order of the
inter-electron spacing). The spin state of the conductor is then static and tunneling events at different
points will  in general result in different spin orderings, as depicted in Fig.\ \ref{fig1}. Interference will only be
possible if all $N$ electrons in the interference loop have the same
spin state. This occurs with the probability $2^{-N}$ and it becomes very unlikely at large $N$. 
A magnetic field with a Zeeman energy $E_Z\gg kT$ large enough to
polarize the electron spins  restores the interference. This loss of coherence at arbitrarily low temperatures contrasts with the decay of persistent currents in isolated rings in the same regime that occurs only at finite temperatures $kT \gtrsim v/lN$ \cite{kn:viefers2004}.

\begin{figure}
\vspace{-2cm}
\includegraphics[width=\hsize]{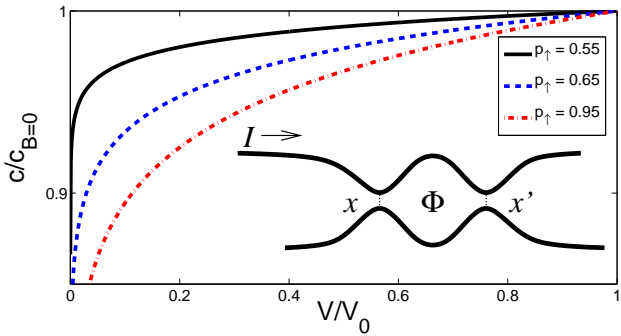}
\vspace{-1cm} \caption{In our second geometry (inset) the interference contrast $C$ obeys a powerlaw with a temperature and magnetic field dependent exponent $g - g\ln^2 p_\uparrow/\pi^2$. We show the normalized interference contrast $c=C(V)/C(V_0)$ compared to its value without magnetic field  $c_{B=0}$,  for various  $p_\uparrow$  ($g=1$).    }  \label{fig2} 
\end{figure}

More formally,
an interference experiment in the geometry of Fig.\ \ref{fig1} probes  the
amplitude for an electron to propagate from point $x$ to $x'$ via different wires,
and thus probes the exponential decay of the single-electron Green function 
with distance  predicted for the spin-incoherent 
regime \cite{kn:cheianov2004,kn:fiete2004}. This decay can alternatively be 
observed in momentum resolved tunneling \cite{kn:auslaender2002,kn:fiete2005}.
It is not a feature of the spin-incoherent regime $l_\varphi^\sigma \lesssim a$ alone, but 
occurs whenever temperatures are large enough that $l_\varphi^\sigma <l$ \cite{LeHur05}.  We show
that in an alternative geometry (shown in Fig.\ \ref{fig2}) the scaling of interference effects with voltage
may be used to distinguish the true spin-incoherent regime $l_\varphi^\sigma \lesssim a$ from the  regime $l_\varphi^\sigma \lesssim l$.  Because of space limitations 
we present explicit calculations only in the spin-incoherent limit $l_\varphi^\sigma \lesssim a$, but we will compare with the
other limit $a \lesssim l_\varphi^\sigma \lesssim l$  at
appropriate points.


We consider two interference geometries: one geometry where electrons tunnel between semi-infinite wires and one where they tunnel between effectively infinite wires, as shown in Figs.\ \ref{fig1}  and \ref{fig2}.  Following Refs.\ \cite{kn:fiete2004,kn:fiete2005} we model a  spin-incoherent wire by a spinless Luttinger liquid
and a static spin background,
 \bea \label{Hincoh}
 H&=&H_{1}+H_{2}+H_T,  \nonumber \\
  H_{\alpha} &=& v \int_0^\infty{\frac{dx}{2\pi}\left[ g^{-1}
     (\partial_x\theta_\alpha)^2+g(\partial_x\phi_\alpha)^2\right]}, \nonumber \\ 
H_T&=&e^{ieV\tau}\sum_\sigma \left[ t \, \psi^\dag_{1\sigma}(x_{1}) \psi_{2\sigma}(x_{2})\right. \nonumber \\
 && \left. \mbox{} \;\;\;\;\;\;+ 
  e^{i\varphi} t'  \psi^\dag_{1\sigma}(x_{1}') \psi_{2\sigma}(x_{2}') 
  + \mbox{h.c.}\right].
\eea
Here, $\alpha = 1,2$, $V$ is the bias voltage, $\tau$ the time argument and $t$ and $t'$ are the 
tunneling amplitudes. 
 We have chosen units such that 
$\hbar=1$. The boson fields obey the standard commutation relation
$[\theta_\alpha(x),\phi_\alpha(x')]=-i \pi\, \Theta(x-x')$. The velocity $v$ is a property of the
spinless charge carriers  $c_\alpha$. The
electron fields $\psi_{\alpha\sigma}$ are expressed in terms of these fermions $c_\alpha$  and operators $S_{\alpha\sigma}(x)$ that add a spin 
$\sigma$ to the spin background of wire $\alpha$ at position $x$ as $ \psi_{\alpha\sigma}(x)=c^\dag_\alpha(x) S_{\alpha\sigma}(x)$ with
\beq \label{field}
c_\alpha(x)=\frac{\eta_\alpha}{\sqrt{2\pi a}}\sum_{n=\pm1} e^{in[\theta_\alpha(x)+ k_Fx]} e^{i\phi_\alpha(x)} 
\eeq
in bosonized form. Here, $\eta_\alpha$ are Majorana fermions and $k_F$ is the Fermi
wavevector of the fermions $c_\alpha$.
The Hamiltonian (\ref{Hincoh})
can be derived microscopically as the low-energy theory of a Hubbard
model with infinite on-site repulsion $U$ and an additional 
long-range interaction described by a purely forward scattering 
density-density coupling. An infinite-$U$
Hubbard model   is described in terms of spinless holes that become the   fermions $c_\alpha$ introduced above and a static spin background  \cite{kn:Bernasconi1975}.   

The total tunnel current $I=I_{\rm dir}+I_{\rm int}$ is the sum of a $\varphi$-independent
contribution $I_{\rm dir}$ and a $\varphi$-dependent interference
contribution $I_{\rm int}$. The two contributions are expressed in
terms of the electron Green functions as
\begin{widetext}
\bea
  I_{\rm dir}&=& 2 |t|^2 \sum_{\sigma} \int d\tau\, e^{-i e V \tau}
  \left[ G^<_{1\sigma}(x,x,\tau)G^>_{2\sigma}(x,x,-\tau)-
  G^>_{1\sigma}(x,x,\tau)G^<_{2\sigma}(x,x,-\tau)\right]
  +
  (\mbox{$t \to t'$, $x \to x'$}) ,
  \label{eq:Idir} \\
  I_{\rm int}&=&2 t^* t' e^{i\varphi}
  \sum_\sigma \int d
  \tau\,e^{-ieV\tau}\left[G^<_{1\sigma}(x,x',\tau)G^>_{2\sigma}(x',x,-\tau)-
  G^>_{1\sigma}(x,x',\tau)G^<_{2\sigma}(x',x,-\tau)\right]
  + \mbox {c.c.} , \label{eq:Iint}
\eea
\end{widetext}
 \begin{eqnarray}
  \label{eq:Ggreater}
  G_{\alpha\sigma}^>(x,x',\tau) &=& -i \langle
  \psi^\pdag_{\alpha\sigma}(x,\tau)\psi^\dag_{\alpha\sigma}(x',0)\rangle
  \\ &=&
  -i\langle S^\pdag_{\alpha\sigma}(x,\tau)c_\alpha^\dag(x,\tau)
  c_\alpha(x',0)S^\dag_{\alpha\sigma}(x',0)\rangle, \nonumber
\end{eqnarray}
and $G^<$ is defined correspondingly.  The spin expectation value in Eq.\ (\ref{eq:Ggreater}) is
non-vanishing only if adding a spin $\sigma$ at position $x'$ and time
$0$ and removing a spin of the same orientation at position $x$
and time $\tau$ does not alter the spin background. Since the spin background in our model of impenetrable electrons follows moving charges,   this occurs with
probability $p_\sigma^{|N_{ x\alpha}(\tau)-N_{x'\alpha}(0)|}$, where $N_{x\alpha}$ is the
 number of electrons to the left of point $x$  and 
\beq
p_\uparrow=1-p_\downarrow=\frac{1}{1+\exp(-E_Z/kT)}
\label{pdef}
\eeq
is the probability for a spin to point along the
direction of the applied magnetic field. Consequently the spin and the charge expectation values in Eq.\ (\ref{eq:Ggreater}) do not factorize as in the case of spin-coherent Luttinger liquids. This underlies most of the phenomena that we discuss in this Letter. We thus obtain \cite{kn:fiete2004,kn:fiete2005}
\bea \label{G}
 G_{\alpha\sigma}^>(x,x',\tau) &=& -i\sum_k  p_\sigma^{|k|} \frac{d\xi}{2\pi}\,e^{i\xi k}    \\
 &&\times \langle e^{-i \xi N_{x\alpha}(\tau)} c_\alpha^\dag(x,\tau) c_\alpha(x',0)e^{i\xi N_{x'\alpha}(0)}\rangle .\nonumber
\eea
$G^<$ is calculated similarly.
After bosonization  one has $N_{x\alpha}(\tau)=[k_F
x+\theta_\alpha(x,\tau)]/\pi$. In this representation, the discrete
nature of $N_{x\alpha}$ is lost, so that our results will only be
approximate. In keeping with the continuous nature of the bosonized
theory, we replace the summation over $k$ in Eq.\ (\ref{G}) by
an integral. With this replacement, our calculation using bosonization approaches the exact results in the spin-polarized limit $p_\sigma 
\to 1$, when the integration over $k$ enforces $\xi=0$ in Eq.\ (\ref{G}).

In order to prevent the interference current $I_{\rm int}$ to be
smeared out by the applied bias voltage, interference experiments need
to be done with a bias voltage $e V \ll v/l$. Therefore, all results
presented in this Letter are for the limit $v \tau
\gg l$. For definiteness, we also assume $p_{\uparrow} \ge
p_{\downarrow}$.

We first analyze the geometry of Fig.\ \ref{fig1}, 
where tunneling occurs between the bulk of one wire and the end of the
other wire. We evaluate the Green functions Eq.\
(\ref{G})   in semi-infinite wires
following Ref.\ \cite{kn:fabrizio1995}.  We give the Green functions of the source wire, $\alpha=1$; Green functions for $\alpha=2$ are obtained by
interchanging the coordinates $x$ and $x'$. For temperatures 
$k T \ll \tau^{-1}$ we find
\bea
  G_{1\sigma}^>(x,x',\tau) & \sim &
   \int dk\, \frac{l^{1/4g}\sin (k_F \delta x)
  \sin (\pi k+ k_F \delta x)}{w (i v \tau + a)^{1/g}}
  \nonumber \\
  && \mbox{}\times p_\sigma^{|k-1/4|} 
  e^{-(k-\langle N_{l}\rangle)^2/w^2},
%
\label{Gint} 
\eea
where $\delta x \ll a$ 
is the distance of $x'$ to the end of the wire, $l\gg a$ is the distance 
between the tunneling points $x$ and $x'$ measured along the wire,
$\langle N_l \rangle = k_F l/\pi$, and we abbreviated 
$w = [g \ln (2  l/a)]^{1/2}/\pi$. Equation (\ref{Gint}) and all other
expressions for Green functions below are up to a numerical
proportionality factor that depends on the high-energy cutoff $a$.
For $\langle N_l \rangle \gg {\rm max}\{w,w^2 
|\ln p_\sigma |\}$, a condition that is fulfilled for 
repulsive interactions if $\langle N_l \rangle \gg 1$, $|k|$ in Eq.\
(\ref{Gint}) can be replaced by $k$. We then
perform the integration over $k$ and find
\bea \label{Gsum}
  G_{1\sigma}^>(x,x',\tau) &\sim& p_\sigma^{\langle N_l\rangle-1/4}\, \frac{
   \langle N_l\rangle ^{1/4g -g/4+
  g \ln^2p_\sigma/4\pi^2}}{ (iv \tau
  + a)^{1/g}}
  \\ && \mbox{} \times \sin ( k_F \delta x) 
  \sin (k_F l + \pi w^2 \ln p_\sigma/2).
  \nonumber
\eea
To evaluate $I_{\rm dir}$  we need two more Green functions,
\bea
  G_{1\sigma}^>(x,x,\tau) &\sim& \int dk\, \frac{ l^{1/2g}
  p_\sigma^{|k|}\cos \pi k \,e^{-k^2/2w^2}}{ w (iv\tau+a)^{1/g}}, \\
  G_{1\sigma}^>(x',x',\tau) &\sim& \frac{\sin^2 (k_F
\delta x)}{(iv \tau +a)^{1/g}}.
\eea

Upon substitution into Eqs.\ (\ref{eq:Idir}) and
(\ref{eq:Iint}), we conclude that at $kT \ll eV$ $I_{\rm dir}$ and $I_{\rm int}$ have
the same voltage dependence, $I \propto (eV)^{2/g-1}$. The
interference current $I_{\rm int}$ is, however, reduced 
relative to $I_{\rm dir}$ through the incoherence of the electron
spins. The interference contrast $C$ depends 
on the average number $N=2\langle  N_l\rangle$ of electrons inside 
the interference loop,
\beq \label{cont}
 C \sim p_\uparrow^{  N} N^{-g/2+g \ln^2p_\uparrow/2\pi^2},
\eeq
where we assumed $(p_\downarrow/p_\uparrow)^N \ll 1$. 
For temperatures $k T \gg e V$, one has $I_{\rm dir} \sim I_{\rm int} 
\sim eV (kT)^{2/g-2}$ while  the interference contrast is still given
by Eq.\ (\ref{cont}).

The first factor  in Eq.\
(\ref{cont}) describes the exponential dependence of the interference
contrast on $N$ that has been anticipated by the argument in the
introduction ($p_\uparrow=p_\downarrow=1/2$ in the absence of a
magnetic field). It leads to an exponential suppression of $C$ at
$E_Z \lesssim k T \ln N$ (for large $N$). 
This remarkable magnetic field dependence that is absent in the spin-polarized case
can be distinguished from the effect of
magnetic
impurities by its strong dependence on the electron density in the
wire via $N$ (tunable by a gate voltage).
In the regime of $a \lesssim l_\varphi^\sigma \lesssim l$ an exponential suppression of $C$ similar to that of Eq.\ (\ref{cont}) occurs. Experimentally it can be distinguished from the spin-incoherent regime  ($kT \gg J$) by the field scale required to polarize the electron gas and thus to restore interference which is raised to $E_Z \approx J$.

The second factor in Eq.\
(\ref{cont}) adds a power law dependence of $C$ on $N$
that is due to quantum fluctuations of $N$. 
A similar scaling with the distance from a boundary has been found in a different context in Ref.\  \cite{kn:Eggert1995}, but the
scaling exponent found here depends on the applied magnetic field and the
temperature. While this anomalous scaling is masked in the geometry of
Fig.\ \ref{fig1} by the much stronger exponential dependence
$p_\uparrow^{N}$, it becomes observable in the geometry of Fig.\
\ref{fig2}, which we now discuss.
 

In the geometry shown in the inset of Fig.\ \ref{fig2}, tunneling
takes place between the bulk regions of both wires, with the distance
between the tunneling points $x$, $x'$ and the ends of the wires being much larger than $v/e
V$. Calculating the Green functions as before, we find 
\begin{eqnarray}
  \label{eq:G1}
  G_{1\sigma}^{>}(x,x,\tau) &=&
  G_{1\sigma}^{>}(x',x',\tau) \\ &\sim& \mbox{}
  \int dk \frac{p_{\sigma}^{|k|} \cos\pi k\, e^{-  k^2/2w'^2}}
  {w' (i v \tau + a)^{1/2 g}}, \nonumber
\end{eqnarray}
where  $w' = [ g \ln (i v \tau/a)]^{1/2}/\pi$, and
\begin{equation}
  G_{1\sigma}^{>}(x,x',\tau) \sim
  p_{\sigma}^{\langle N_l \rangle}
  \frac{\cos(k_F l + \pi w'^2 \ln p_\sigma)}{(i v \tau + a)^{g/2 + 1/2g
  - g \ln^2 p_{\sigma}/2 \pi^2}}.
  \label{eq:G2}
\end{equation}
In the derivation of Eqs.\ (\ref{eq:G1}) and (\ref{eq:G2}) we again
assumed $\langle N_l \rangle \gg 
\max[|w'|,|w'|^2 |\ln p_{\sigma}|]$. Equations (\ref{eq:G1}) and
(\ref{eq:G2}) reduce to the Green functions obtained in Ref.\
\cite{kn:fiete2004} in the limit of zero magnetic
field, $p_{\sigma} = 1/2$. For $g=1$, when describing an infinite-$U$ Hubbard model,
they moreover coincide with the Green functions found in Ref.\ \cite{kn:cheianov2004} 
without bosonization and the resulting loss of the discreteness of charge. 
For not too large magnetic fields, $E_Z\ll g kT \ln (v\tau/2a)$, 
the $k$-integral in Eq.\ (\ref{eq:G1}) is to a good approximation $\tau$-independent. 
The resulting voltage dependence of the tunneling current is different than for
the geometry of Fig.\ \ref{fig1}: we find $I_{\rm dir} \propto (e
V)^{1/g - 1}$, while the bias dependence of $I_{\rm int}$ exhibits the
anomalous scaling observed in the length dependence of Eq.\
(\ref{cont}),
\beq
 I_{\rm int} \sim  p_\uparrow^{N}(eV)^{g+1/g - g\ln^2 p_\uparrow/\pi^2-1},
\eeq
where, again,   $p_{\uparrow} \ge 1/2$.
The scaling
of $I_{\rm int}$ with $eV$ is  temperature and magnetic field dependent through $p_\uparrow$,
Eq.\ (\ref{pdef}). 
This is a defining signature of 
spin-incoherence in quantum wires and in particular serves to distinguish the truly spin-incoherent limit $kT \gg J$ from the regime   $a\lesssim l_\varphi^\sigma \lesssim l$. The interference
contrast in this geometry acquires a voltage dependence
\beq
 C \sim  p_\uparrow^{N} (eV)^{g - g\ln^2 p_\uparrow/\pi^2}.
\eeq
\begin{figure}
\vspace{-5.2cm} \hspace{2.5cm}
\includegraphics[width=5cm]{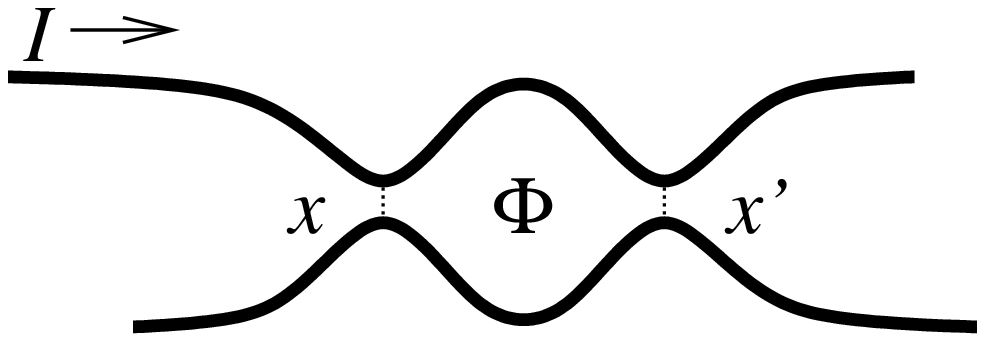}
\vspace{2.7cm}
\end{figure}
Also such scaling of the interference contrast with the applied voltage 
is absent in spin-polarized conductors. The interference contrast $C$ crosses over to the spin-polarized behavior at large magnetic fields $E_Z\gtrsim g kT \ln (v\tau/2a)$ through the $\tau$-dependence of the $k$-integral in Eq.\ (\ref{eq:G1}).
 
Our calculation has been done using equilibrium Green functions in (semi-)infinite wires. It applies to wires of
finite length $L$ if $e V \gg v/L$. The equilibrium assumption
is justified for  spin relaxation times $\tau_{\rm s}$
in the wire that are shorter than the transit time, $\tau_{\rm s} \ll k_F L
e/I$.


In conclusion, we have proposed interference experiments for distinguishing the spin-polarized from the spin-incoherent regime of strongly interacting wires. We
identified two unique signatures of spin-incoherence: A strong 
dependence of the interference contrast on an applied magnetic field
and a power law dependence of the interference contrast on the applied
voltage, if the tunneling takes place between bulk regions of the
quantum wires. The scaling exponent of the interference current is 
surprisingly temperature and magnetic field dependent, another feature
that is unknown from spin-polarized Luttinger liquids. These clear and
qualitative signatures of spin-incoherence make interference
experiments promising tools in the search for the mechanism of the
observed conductance anomalies in interacting one-dimensional
conductors.

This work was supported by the NSF under grants no. DMR 0431350 (AM) and  DMR 0334499 and 
by the Packard Foundation (PB and MK).

  \end{document}